# HIGHER ORDER MODES IN THIRD HARMONIC CAVITIES AT FLASH

I.R.R. Shinton[†*], N. Baboi[‡], T. Flisgen[¥], H.W. Glock[¥], R.M. Jones[†*], U. van Rienen[¥], P. Zhang[‡*]

[†]School of Physics and Astronomy, The University of Manchester, Manchester, U.K.
[*]The Cockcroft Institute of Accelerator Science and Technology, Daresbury, U.K.
[‡]DESY, Hamburg, Germany.
[¥]Universität Rostock, Rostock, Germany.

*Abstract*

Transverse modes in the 3.9 GHz cavities, designed and fabricated by FNAL, are reported on. These modes have the potential to cause significant emittance dilution if they are not sufficiently suppressed. Recent probe-based-experiments have indicated significant discrepancies between modes predicted in stand-alone 9-cell cavities compared to those in 4-cavity modules. We employ the code HFSS [1] to analyze these modes, coupled through beam tubes whose cut-off is above that of the first dipole band.

## INTRODUCTION

Recently a third harmonic module, consisting of a four (nine-cell) cavity string operating at 3.9GHz, has been incorporated into the FLASH facility at DESY [2]. This module is positioned between ACC1 and the bunch compressor. The purpose of this 3.9GHz module is to linearize the energy distribution along the bunch before the bunch compressor. This is of particular importance to current and future accelerators such as XFEL and the ILC. The 3.9GHz structures have significantly smaller dimensions than their 1.3GHz (TESLA cavity) counter parts. Hence the wakefields within the third harmonic module will be considerably larger and more sensitive to fabrication errors. It is therefore important to ensure that these wakefields are adequately damped. A further concern, which was observed from both experiment and simulation [2], is the large number of multicavity higher order modes (HOM's) that are present and propagate throughout the entire 3.9GHz module. A concerted experimental (detailed in next section) and simulation study is in progress to instrument the 3.9GHz cavities in order to characterise these modes and to instrument the cavities; the aim of which is the implementation of a HOM BPM system. This will enable alignment of the beam to the electrical centre of the cavity and to ascertain internal misalignments.

In order to understand the physics behind the HOM's in the 3.9GHz cavity, especially the interactions of the multicavity modes, the 4 cavity string must be simulated in its entirety. This is computationally very expensive in terms of time and resources. In order to facilitate a rapid generation of results, including the effects of fabrication tolerances, we apply a generalised scattering matrix (GSM) approach [3]. This is a well established RF technique that enables the rapid calculations of large structures through the discretization of the entire structure into a series of smaller structures that can be simulated separately and then concatenated using matrix techniques to obtain the complete structure.

A similar technique, the coupled scattering calculation (CSC) has previously been employed [2] to model the entire 3.9GHz module. In this paper we apply similar analysis using the GSM technique to this structure described in the foregoing sections. The following section describes experimental measurements; this is preceded by the simulation of the full cavity string, but prior to this benchmarking of the GSM procedure is outlined. Finally this paper is concluded with final remarks.

## MEASUREMENTS OF HOMS AT FLASH

ACC39 is composed of four interconnected cavities, referred to as C1 through C4. There are two distinct coupler designs: a 1-leg-coupler design (C1 and C3) and a 2-leg-coupler design (C2 and C4), this is illustrated in Fig. 1. Probe based measurements across the string of cavities within ACC39 were conducted using a Rohde and Schwarz-ZVA8 vector network analyser taken at the ACC39 HOM board rack and scanned over a frequency range of 3.5 GHz to 8 GHz with a step of 10 kHz. Additional details of the experimental setup and mode spectra measurement of the ACC39 module after its installation at the FLASH facility in DESY are described in [2]

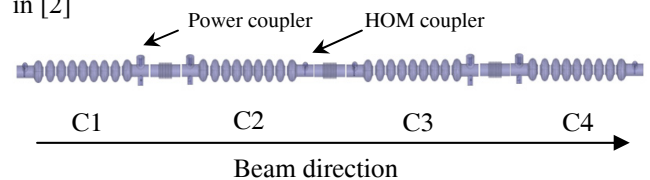

Figure 1: Schematic representation of 4 cavities within ACC39 [2]. The power couplers are indicated in green. C1 and C3 contain the 1-leg-HOM couplers whilst C2 and C4 contain the 2-leg-HOM couplers.

## GSM UNIT CELL METHODOLGY

There are two types of couplers in ACC39 (illustrated in Fig 1). In applying the GSM method, 6 main structures must be simulated (including the two couplers). In addition two input cells are needed to drive power through the structure, making a total of 8 structure types. These cells are illustrated in Fig 2.

All the multi-moded 2-port (single cell and beam pipe), 3-port (end cell with HOM-coupler) or 4-port (endcell with HOM power coupler) scattering matrices used in the

GSM method were generated using the driven modal solver of HFSSv11.2.1. Since the excitation field polarisation of a mode on a wave port in HFSS is calculated in an arbitrary direction, it is necessary to align specific modes. This is of particular importance during the cascade calculation in which the propagation of specific modes must be maintained throughout the calculation. This is achieved within HFSS by aligning specific modes with a calibration direction, which are universally applied (with respect to the chosen coordinate system) to all unit cells. An example of these calibration lines is demonstrated in Fig 3.

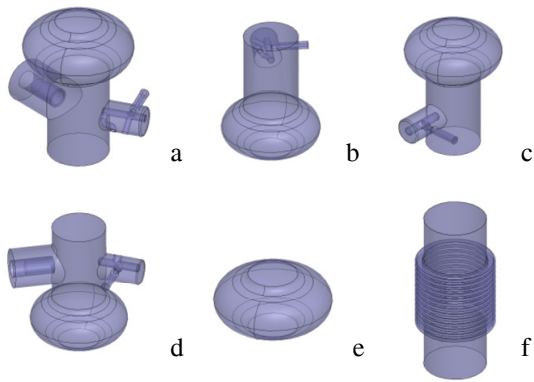

Figure 2: Unit cells used in the cascade calculations of the FNAL ACC39 Cryo-module.

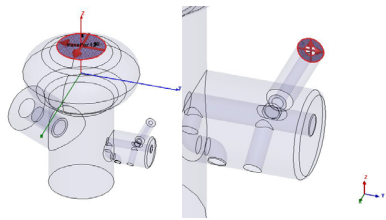

Figure 3: Polarisation directions used in the unit cells within HFSSv11.2.1

Furthermore in the normalisation process it is important to note that HFSS has the facility to re-normalise the port impedance to a specific impedance value (denoted by the user). This has the effect of removing the frequency dependency from the port impedances. This is implemented within HFSS using:

$$Z = \sqrt{Z_o}(I-S)^{-1}(I+S)\sqrt{Z_o} \quad (1)$$

in which $S$ is the $n \times n$ generalized Scattering matrix, $I$ is an $n \times n$ identity matrix and $Z_o$ is a diagonal matrix having the characteristic impedance ($Z_o$) of each port as a diagonal value. The Scattering matrix is readily obtained from Eq (1) re-normalised to the port impedance.

$$S_\Omega = \sqrt{Y_\Omega}(Z-Z_\Omega)(Z+Z_\Omega)\sqrt{Z_\Omega} \quad (2)$$

In Eq (2) $Z$ is the structure's unique impedance matrix and $Z_\Omega$ and $Y_\Omega$ are diagonal matrices with the desired impedance and admittance as diagonal values.

All unit cell simulations used in the GSM calculations were conducted using HFSSv11.2.1, in which a seed mesh, consisting of a surface mesh of 30um, with an aspect ratio of 3, and a volume mesh of 10000 elements was used. The mesh was adaptively refined until a convergence criteria of less than 0.01% was achieved, at which point a fast frequency sweep (using an ALPS algorithm) was made at the upper limit of the frequency range of interest with 1MHz frequency steps, and with 10 modes per port. A fast frequency sweep was chosen over the more robust (albeit slower) discrete frequency sweep as it accelerates the generation of results by a factor of 10 or more. Due to the inherent stability requirements of the ALPS algorithm, all simulations were cross-checked against a discrete frequency sweep (in which a larger frequency step was used). We were primarily interested in investigating the first two dipole bands regions and hence two frequency ranges were specified: 4GHz to 5GHz and 5GHz to 6GHz.

To demonstrate the validity of the GSM method we present a benchmarking problem in which the GSM calculation is compared to a full structure simulation in HFSS. This is shown below in Fig 4 in which we propagate modes from the HOM port of C1 through to the iris of a middle cell. We observe excellent agreement between the GSM method and the full scale simulation. This gives some confidence in the GSM method and allows progression on to simulating the 3.9GHz module in its entirety; the results of which are presented in the next section.

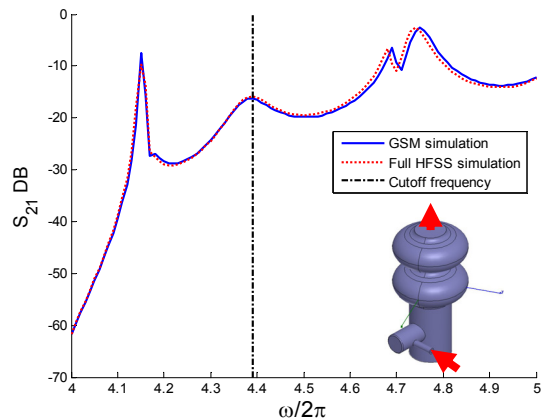

Figure 4: Benchmarking comparison of one TE11 polarisation between a full HFSSv11.2.1 simulation and that calculated by GSM. The simulated structure is shown insert. The dashed black line represents the beam-pipe cut-off.

It is important to bear in mind that the GSM approach requires at least a single mode to be propagating within the input and output ports [3]. TEM modes clearly satisfy this criteria.

## MEASUREMENTS AND SIMULATION OF MODE SPECTRA

Results of applying the GSM method across the full ACC39 cavity string (from the first HOM port of C1 to the last HOM port of C4) in range of both the first and second dipole bands are presented below in Fig 5.

Note that as we are using the GSM method of [3], effectively a 2-port cascading method, we are unable to calculate the effect of multicavity modes measured from the ports of central cavities as this would facilitate a 3 port unit cell calculation (refer to Fig 6). Hence a direct comparison with the multicavity N-port CSC results of [2] are possible with this particular GSM method if the full ACC39 cavity string (depicted in Fig 6) is simulated.

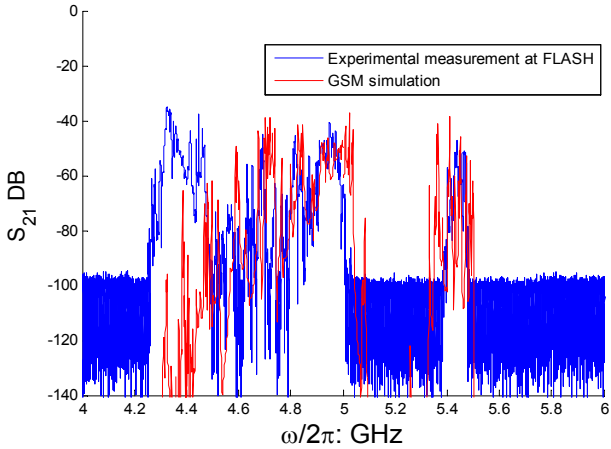

Figure 5: Direct comparison GSM with the experimental data taken across the entire ACC39 string for transmission from C1, upstream HOM-coupler, to C4, downstream HOM-coupler. The GSM S matrices have been re-normalised.

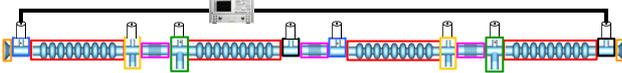

Figure 6: Diagrammatic representation of GSM structural calculation.

Below 4.39GHz the agreement between simulation and experiment is poor. We attribute the discrepancy as being due to modes below cut-off (the beam-pipe cut-off is ~4.39GHz for dipole modes) in the cascaded blocks. The method relies on there being at least one propagating mode present in order to be accurate.

Detailed investigations on the measured spectra indicate that the modes above the cut-off frequency of the beam tubes are shifted in frequency with respect to their isolated cavity values. This is a possible indication of multi-cavity modes being present in the module. The frequencies measured in ACC39 include both coupling of cavity modes and frequency shifts due to fabrication errors.

From our GSM simulations in which we compare single cavity results (of C1 removed from the cavity string) to those of a cavity string (consisting of conjoined C1 and C2 cavities) as seen in Fig 7 we observe a shifting of the modal frequencies above the cut-off frequency of the beam tubes (~ 4.39 GHz for dipole modes) with respect to their isolated cavity values. This clarifies the strong influence of inter-cavity coupling.

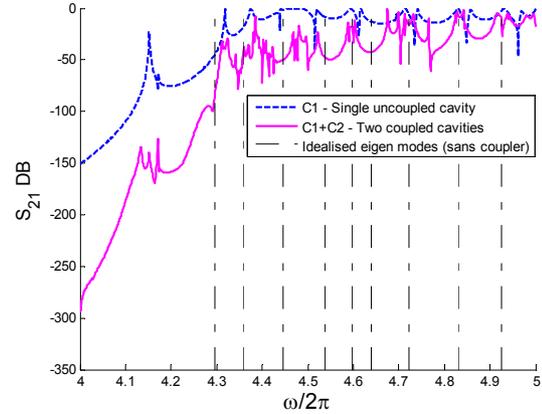

Figure 7: Comparison of the GSM results calculated for the C1 cavity (single cavity alone) to that of a concatenated C1+C2 (two coupled) cavity strings for the first dipole band. The green dashed vertical lines represent the resonant eigenmodes calculated for an idealised cavity (without couplers) [4].

## FINAL REMARKS

From our GSM coupled cavity simulations we observe splitting of the modes as a result of the couplers and the strong coupling between neighbouring cavities. Interference effects are evident in which there is coupling through the beam-pipes of the dipole modes in adjacent cavities. Finally we note that these results demonstrate a qualitative agreement with the experimental values. Future work will be concerned with an additional characterisation of the dipole modes both with and without beam-excitation, with a view to determining a mode suitable for the electronics for beam and cavity alignment studies.

## ACKNOWLEDGEMENTS

This research has received funding from the European Commission under the FP7 Research Infrastructures grant no. 227579. We are pleased to acknowledge H. Ewald, for providing the R&S-ZVA8-network analyser and D. Mitchell, W.D. Möller and E. Vogel [5] for contributing important geometrical details on the cavities and couplers within ACC39.